\documentstyle[preprint]{jpsj}

\def\runtitle { 
A Multicanonical Molecular Dynamics Study on a Simple Bead-Spring Model for Protein Folding
}

\def\runauthor{Masaharu {\sc Isobe}, Hisashi {\sc Shimizu} and Yasuaki {\sc Hiwatari}}

\title { 
A Multicanonical Molecular Dynamics Study on a Simple Bead-Spring Model for Protein Folding
}

\author
{ 
Masaharu {\sc Isobe}$^{1,3}$\footnote{E-mail: isobe@cphys.s.kanazawa-u.ac.jp},
 Hisashi {\sc Shimizu}$^{2}$ and Yasuaki {\sc Hiwatari}$^{3}$
}

\inst
{
$^1$Japan Science and Technology Corporation (JST) \\
$^2$Department of Physics, Shinshu University, 3-1-1 Asahi, Matsumoto, 390-8621 \\
$^3$Department of Computational Science, Kanazawa University, Kakuma, Kanazawa, 
920-1192
}

\recdate
{
\today
}

\abst{ 
We have performed a multicanonical molecular dynamics simulation on a simple model protein.
We have studied a model protein composed of charged, hydrophobic, and neutral spherical bead monomers.
Since the hydrophobic interaction is considered to significantly affect protein folding, we particularly focus on the competition between effects of the Coulomb interaction and the hydrophobic interaction.
We found that the transition which occurs upon decreasing the temperature is markedly affected by the change in both parameters and forms of the hydrophobic potential function, and the transition changes from first order to second order, when the Coulomb interaction becomes weaker.
}

\kword{ 
multicanonical molecular dynamics simulation, continuum model, bead-spring model, protein folding, solvent effect, Coulomb interaction, hydrophobic interaction 
}

\begin{document}
\sloppy
\maketitle

Protein structures and folding mechanisms are mysterious subjects.\cite{pande_2000}
According to Anfinsen's dogma~\cite{anfinsen_1973}, all of the information needed to construct a protein's three-dimensional structure is contained within its amino acid sequence.
For given interactions between amino acids, we should in principle obtain a unique folding solution for transformation of proteins to their native states.
However, from a numerical point of view, the prediction of a protein's native structure for a given amino acid sequence is considerably difficult because conformations accessible to a given polypeptide chain grow exponentially with chain length and it would require a much longer time than any computationally accessible time scale, known as the Levinthal (Levinthal paradox).\cite{levinthal_1968}

In order to understand on a molecular level the general features of protein structures and mechanism of folding, it is important to know the role of four major molecular interactions related to protein folding, which are hydrogen bonds, and van der Waals, hydrophobic, and Coulomb interactions.
Computational studies for understanding protein folding mechanisms have been performed often under two particular assumptions, using either the simplest protein model with a lattice or all-atom model together with solvent molecules in a continuous system.
With these two extreme models, it is not easy to capture the nature of the intrinsic interaction in protein folding.
To address this problem, one of the most promising methods is to examine simpler protein systems, such as bead-spring models, specifically designed to simulate the role of interactions.

In general, complex systems with some different interactions often show frustration of potential energies, and thus exhibit a huge number of local-minimum-energy states in a free-energy landscape.
Simulations of these systems by conventional molecular dynamics methods tend to get trapped in either of the local-minimum-energy states at low temperatures, and thus a long CPU time is needed to obtain accurate physical distributions since the relaxation time increases markedly with increasing complexity.
Recently, numerical algorithms for overcoming such a multiple-minima problem have been proposed, such as the multicanonical ensemble method~\cite{berg_1992}, which allows the trajectory of particles to escape over energy barriers and consequently enables sampling over a much wider phase space than the conventional method does.

In this paper, we study a simple real protein, protein-g (PDB id: 2gb1), which is composed of 56 amino acids. 
Figure 1 shows the distance map for protein-g, where filled symbols indicate the pairs of residues whose $\alpha$-carbons are at a distance less than 13.0 [\AA].\cite{saito_1999}
Our model for protein-g is composed of positively ($+$) or negatively ($-$) charged, hydrophobic (H), and neutral (P) spherical bead monomers with the following sequences: ${\rm HPP}+{\rm HHHPH}+{\rm PH}+{\rm H}-{\rm PPP}-{\rm HP}-{\rm HHPH}-+{\rm HH}+{\rm PPHP}-{\rm PHH}-{\rm H}-{\rm PPP}--{\rm HP}+{\rm PHPHP}-$.
Using this model for protein-g, we performed a multicanonical molecular dynamics (MMD) simulation,\cite{nakajima_1997,shimizu_1999} systematically. 
The main purpose of the present study is to examine the relative contributions among the respective interactions between monomers using molecular dynamics simulation, which may make it possible to construct the universal framework in protein folding.
Since the hydrophobic interaction is considered to have a large effect on protein folding, it is of interest to observe how the resultant protein structure changes as the hydrophobic interaction sets in.
Therefore, we employ the following interactions between monomers:(i) the excluded volume interaction (soft-core model), (ii) the covalent bond interaction (spring model),  (iii) Coulomb interaction, and (iv) hydrophobic interaction.
Recently, molecular dynamics simulations for the model with (i)-(iii) interactions have been performed by Baumketner et al.~\cite{baumketner_2000}
In the present study, the hydrophobic interaction, which originates from solvent effects, is also involved.
Thus, in this paper we will in particular discuss the competition between effects of the Coulomb interaction and those of the hydrophobic interaction.
Because of the lack of short-range attractive interactions such as hydrogen bond and van der Waals interaction, this model does not reproduce the true native structure of protein-g.
However, since in this paper we mainly address conformational changes of a protein driven purely by long-range interactions, this is beyond the scope of the present study.
Note that most theories of solution dynamics for homopolymers are based on the bead-spring model.\cite{erman_2000,doi_1986}
Advantages of the bead-spring model, compared to other simplified models, such as lattice models, are that the model is practical, finds energy minima with less computation, and the physical properties obtained can be directly related to experiments.

A solvent (like water) generally has the following two effects:(a) because of a screening effect, the interaction between charged monomers becomes significantly weaker (i.e., dielectric constant $\epsilon_c^{*} \gg 1.0$), (b) hydrophobic monomers repel water molecules in their neighborhood to result in a construction of hydrophobic cores and effectively gives rise to an attractive interaction (strength of the hydrophobic interaction is designated with $\epsilon_{hb}$).
Such solvent effects are non-trivial, because the number of associated solvent molecules is very large.
In order to incorporate solvent effects into our model effectively, we tried to assume two different types of pair potential models for the hydrophobic interactions, the power potential $V_{hb}^{pow.}$ and the exponential potential $V_{hb}^{exp.}$, which are given by
\begin{equation}
V_{hb}^{pow.}(r)=4\epsilon_{sc}(\frac{\sigma_{sc}}{r})^{12}-4\epsilon_{hb}(\frac{\sigma_{hb}}{r})^6,
\end{equation}

\begin{equation}
V_{hb}^{exp.}(r)=\{
\begin{array}{lc}
4\epsilon_{sc}(\frac{\sigma_{sc}}{r})^{12}-\epsilon_{hb}\exp{(-\frac{r-\sigma_v}{d})} & (r>\sigma_v), \nonumber \\
4\epsilon_{sc}(\frac{\sigma_{sc}}{r})^{12}-\epsilon_{hb} & (r \leq \sigma_v),
\end{array}
\end{equation}

\noindent
where $\sigma_v$ and $d$ are 5.5 [\AA] and 10.0 [\AA], respectively.
$\epsilon_{sc}$ is fixed at 3.0 [kcal/mol], and both $\sigma_{sc}$ and $\sigma_{hb}$ are fixed at 3.8 [\AA], which is the same value as the bond length between adjacent monomers. 
In the case of the power potential, which was introduced by Shea et al.~\cite{shea_1998}, the interaction reaches only a few neighbors.
On the other hand, in case of the exponential potential, which was used by Israelachvili et al.,~\cite{israelachvili_1982} the attractive force reaches a much longer distance such as Coulomb interactions.
As mentioned above, since the explicit form of the hydrophobic interaction is not yet well known, it is interesting to study the effects of the form of the hydrophobic interaction on the folding process.
Therefore, we study here both types of potentials and compare the results obtained.
In our simulation, we regard $(\epsilon_c^{*},\epsilon_{hb})$ as the control parameters representing the strength of the Coulomb interaction relative to the hydrophobic interaction.
A change of the control parameters corresponds to a change of solvents. 
We carried out various simulations including a two-limit system, i.e., pure vacuum and pure water (Table I).

\begin{table}
\caption{
The change of strength of the Coulomb interaction relative to the hydrophobic interaction in a two-limit system (i.e., pure vacuum and pure water).}
\label{tbl:1}
\begin{tabular}{@{\hspace{\tabcolsep}\extracolsep{\fill}}ccc}\hline
Interaction   & pure vacuum        & pure water \\ \hline
Coulomb       & large ($\epsilon_c^{*} \sim 1$)   & small ($\epsilon_c^{*} \sim 80.0$) \\
Hydrophobic   & small ($\epsilon_{hb} \sim 0.0$)   & large ($\epsilon_{hb} \sim 3.0 {\rm [kcal/mol]}$) \\ 
\hline
\end{tabular}
\end{table} 

To analyze our data obtained by the present MD calculations, we used the following two order parameters, namely the radius of gyration (RG), which shows an order of spatial expansion of polymers quantitatively (eq.~(\ref{rg})), and the distance matrix error (DME), which measures the difference between structures of the model protein and a reference system ${\bf r}_{ij}^C$ (eq.~(\ref{dme})); for the latter, we have chosen the native structure of protein-g:

\begin{equation}
RG=\sqrt{\frac{1}{N}\sum_{i=1}^N {\bf r}_i^2-(\frac{1}{N}\sum_{i=1}^N {\bf r}_i)^2},
\label{rg}
\end{equation}

\begin{equation}
DME=\sqrt{\frac{2}{N(N-1)}\sum_{<i,j>}^N ({\bf r}_{ij}-{\bf r}_{ij}^C)^2},
\label{dme}
\end{equation}

\noindent
where $N$ is the number of residues, and $r_{ij}=|{\bf r}_i-{\bf r}_j|$.

In MMD simulation, forces acting on particles are biased with a factor as shown below in eq.~(\ref{f_mmd}) which is calculated from the probability function of potential energy at each step so that we can obtain the flat distribution of potential energy $P(E)$ after several preliminary simulations (eq.~(\ref{e_mmd})).

\begin{eqnarray}
{\bf {\cal F}}(t) & = & -{\bf \nabla} {\cal E}(E) = -\frac{d {\cal E}(E)}{dE}{\bf \nabla} E, \nonumber \\
              & = & -(1+k_BT_0\frac{d}{dE}\log{P_{T_0}(E)}){\bf \nabla} E,
\label{f_mmd}
\end{eqnarray}

\begin{equation}
{\cal E}^{new}(E)={\cal E}^{old}(E)+k_BT_0\log{P_{T_0}(E)},
\label{e_mmd}
\end{equation}

\noindent
where $k_B$, $T_0$, and $P_{T_0}$ are the Boltzmann constant, a reference temperature and the canonical distribution function at temperature $T_0$, respectively.

In Fig.~2, the probability distribution functions (PDF) of the potential energy during multicanonical runs and canonical runs (100, 200, and 300 [K]) are shown, in which the hydrophobic interaction obeys a power potential and the control parameters are $(\epsilon_c^{*},\epsilon_{hb})=(5.0,3.0)$ at 300 [K]. 
The inset also shows the evolution of the potential energy during MMD simulation (10 million steps).
Note that the time mesh in MD integration is fixed at 1 [fs] throughout this paper.
It can be seen that the PDF through MMD after 30 iterations becomes flat over a wide range of the potential energy.

Figure~3 shows the PDF obtained by MMD runs (and CMD runs in the inset) in two parameter planes, RG and DME, in which the hydrophobic interaction and the control parameter are the same as those in Fig.~2.
It turns out that MMD runs samples more effectively than CMD runs do for the same number of samples (100 million).~\cite{comment1}

Using the reweighting method~\cite{ferrenberg_1988} to the probability distributions in MMD runs, we obtain canonical distributions for a wide range of temperatures. 
Figure 4 shows the temperature dependence of RG obtained in such a way.
For the case that the hydrophobic interaction and the control parameter are the same as those in Fig.~2 (top-right panel of Fig.~4), we find two stable states with different RG values at approximately 300 [K] and a barrier between them, which suggest that the transition between these two states is of the first order.
The corresponding distance maps for these two stable states are shown in Fig.~5.
The right panel in Fig.~5 indicates that the C-terminal of our model protein-g is far from a folded state, in which RG takes a value as large as $\sim$ 10 [\AA] at approximately 300 [K].
When the hydrophobic interaction becomes weaker, for example $(\epsilon_c^{*},\epsilon_{hb})=(5.0,2.0)$ as shown in the top-left panel of Fig.~4, the transition seems to change from the first order to the second order.
On the other hand, no transition exists for the case that the hydrophobic interaction is given by an exponential potential (bottom panels of Fig.~4).
It follows that the existence of the transition changes markedly when the form of hydrophobic interaction is varied.~\cite{comment2}

To understand the influence of frustrations between several potentials more clearly, we employed weak Coulomb interactions to suppress frustrations.
Figure 6 shows the temperature dependence of RG in weak Coulomb interactions, which corresponds to pure water $(\epsilon_c^{*},\epsilon_{hb})=(80.0,3.0)$ as shown in Table 1.
We found that the transition is of the second order instead of the first order at approximately 300 [K] despite the fact that the hydrophobic interaction has the same value as that in the top-right panel of Fig.~4.

Here, let us consider the relationship between the present results and the experimental results for the gel phase transition. Polymers in the gel have only charged residues.\cite{takeoka_1999}
For the gel phase transition to be of the first order, sufficient attractive and repulsive interactions are necessary.
This suggests that the transition changes from the first order to the second order for the case that the Coulomb interaction becomes substantially weak.
Our results for a model protein also show the same tendency for the first-order phase transition to be induced.
In biopolymer systems, both Coulomb and hydrophobic interactions give rise to the energy frustration that plays a crucial role in the folding process.
The character of transition is clearly changed when the solvent is varied.~\cite{comment3}
Our results suggest that the Coulomb interactions, both repulsive and attractive, contribute to the existence of the first-order transition.
The general protein-folding mechanism is not yet known.
Simulations to study the phase diagram for various model protein systems in various solvents are helpful and currently being undertaken by us.

Finally we give below a brief summary of our work.
We have performed multicanonical molecular dynamics simulations on a simple model for protein-g.
Since we particularly focus on the competition between effects of Coulomb and hydrophobic interactions, the control parameters ($\epsilon_c^{*}, \epsilon_{hb}$) are widely varied.
We found that the present model for protein-g exhibits either continuous or discontinuous phase transition, indicating explicit evidence of frustrations usually observed in heteropolymer systems.
RG dependence of temperature for each control parameter leads to the conclusion that the nature of the structural phase transition as a function of temperature is significantly affected by both the strength and the form of the hydrophobic interaction.

The authors thank Dr. Andrij Baumketner for valuable discussions.
This work was supported by the Research and Development for Applying Advanced Computational Science and Technology (ACT-JST).


\begin{figure}
\caption{
Distance map of protein-g (PDB id: 2gb1).
The pairs of residues within 13.0 [\AA] are filled.
The full sequence of the type of residues is also shown.
}
\label{fig.1}
\end{figure}

\begin{figure}
\caption{
The probability distribution functions of the potential energy for multicanonical runs (MMD) together with canonical runs (CMD) at 100, 200, and 300 [K]. The hydrophobic interaction obeys a power potential and the control parameters are $(\epsilon_c^{*},\epsilon_{hb})=(5.0,3.0)$ at 300 [K]. 
The inset also shows the evolution of potential energy during MMD simulation (10 million steps).
}
\label{fig.2}
\end{figure}

\begin{figure}
\caption{
The probability distribution map in RG-DME plane obtained by multicanonical and canonical (inset) runs.
The horizontal axis shows RG in the range between 0 and 20 [\AA], and the vertical axis shows DME in the range between 0 and 20 [\AA].
The hydrophobic parameters are the same as those in Fig.~2.
}
\label{fig.3}
\end{figure}

\begin{figure}
\caption{
The temperature dependencies of the probability distribution of RG are shown for various control parameters.
The top figures are for the power hydrophobic interaction, and the bottom ones are for the exponential hydrophobic interaction.
The horizontal axis shows RG in the range between 0 and 20 [\AA], and the vertical axis shows temperature in the range of 200 $\sim$ 600 [K].
}
\label{fig.4}
\end{figure}

\begin{figure}
\caption{
The distance maps for two stable states corresponding to the top-right panel of Fig.~4 at approximately 300 [K].
}
\label{fig.5}
\end{figure}

\begin{figure}
\caption{
The temperature dependence of the probability distribution of RG in pure water is shown.
The horizontal axis shows RG in the range of 0 $\sim$ 20 [\AA], and the vertical axis shows temperature in the range of 200 $\sim$ 600 [K].
}
\label{fig.6}
\end{figure}


\begin{thebibliography}{99}
\bibitem{comment1}
In Fig.~3, it turns out that this model protein has a minimum free energy in the region of the localized high probability.
When the hydrophobic interaction becomes weaker, the region of the high probability extends over in RG-DME plane. This indicates that the conformation of minimum free energy is substantially degenerated.
\bibitem{comment2}
We obtained the same temperature dependence for DME as for RG. Both show the same tendency about the nature of the transition, i.e., first order or second order.
Therefore, the results for RG are shown here.
\bibitem{comment3}
To our knowledge, experiments on protein folding in various solvents (such as ethanol) have not been studied yet.
Our simulation must be useful in understanding the role of long-range interactions in an early stage of protein folding in different solutions.
\end{thebibliography}
\end{document}